\documentclass[twocolumn,prl,aps]{revtex4}

\usepackage[dvips]{graphicx}
\usepackage[dvips]{graphics}

\begin{document}

\title{Quantized spin waves in the metallic state of magnetoresistive manganites}

\author{S. Petit$^1$, M. Hennion$^1$, F. Moussa$^1$, 
D. Lamago$^{1,2}$ A. Ivanov$^3$, Y. M. Mukovskii$^4$, D. Shulyatev$^4$}

\affiliation{$^1$ Laboratoire L\'eon Brillouin, CEA-CNRS, CE-SACLAY, F-91191 Gif sur Yvette Cedex, France \\
$^2$ Forschungzentrum Karlsruhe, INFP, P. O. Box 3640, D-76021 Karlsruhe, Germany\\
$^3$ Institut Laue Langevin, 156X, 38042 Grenoble cedex 9, France \\
$^4$ Moscow State Steel and Alloys Institute, Moscow 119049, Russia}
\date{\today}

\begin{abstract} 
High resolution spin wave measurements have been carried out in ferromagnetic (F) $La_{1-x}(Sr,Ca)_xMnO_3$ with $x(Sr)$=0.15, 0.175, 0.2, 0.3 and $x(Ca)$=0.3. In all q-directions, close to the 
zone boundary, the spin wave spectra consist of several energy levels, with the same values in the metallic and the x$\approx$1/8 doping ranges. Mainly the intensity varies, jumping from the lower energy levels determined in the x$\approx$1/8 range to the higher energy ones observed in the metallic state. On the basis of a quantitative agreement found for $x(Sr)=0.15$ in a model of ordered $2D$ clusters, the spin wave anomalies of the metallic state can be interpreted in terms of quantized spin waves within the same $\it 2D$ F clusters, embedded in a {\it 3D} F matrix.
\end{abstract}


\maketitle
Charge segregation is likely the most fruitful concept to understand 
the extraordinary properties of manganites \cite{Dagotto}.
In spite of more than one decade of studies with recent 
theoretical improvements \cite{Motome,Kumar}, the doping driven transition 
from the orbitally ordered ({\it OO}) insulating $LaMnO_3$ to the orbital 
disordered ({\it OD}) metallic phase and the true nature of this metallic 
ferromagnetic (F) phase in cubic manganites remain very puzzling. Most of 
previous works have reported strong anomalies of the spin waves specially 
close to the zone boundary\cite{Endoh,Ye,Moussa2}. Their importance comes 
from their general character, so that they are thought to be generic 
features of the metallic 
state.

We have performed high resolution measurements of spin wave excitations 
in five $La_{1-x}B_xMnO_3$ samples, $B=Sr$ or $Ca$, for dopings covering a large part of the 
phase diagram (see Figure 1), from the quasimetallic (zone 2) and insulating 
(zone 3) parts with $x(Sr)=0.15$ to the metallic part with $x(Sr)=0.175, 
0.2, 0.3$ and $x(Ca)=0.3$ (zone 4). In all of them, the magnetic excitation 
spectrum consists of a quadratic dispersed curve in the zone center, 
characteristic of a three-dimensional ($3D$) ferromagnetic state, and wavevector-independent 
levels in the zone boundary. The very new result is that the energy 
levels observed in the metallic state have the same values whatever the doping 
content (zones 2, 3, 4), the temperature $T$ or the average cation size $r_a$\cite{Hwang}. The intensity 
of each level depends however on $x$, $T$ or $r_a$ and is found to jump from the 
lower ones ($\approx$ 15, 22 meV along [100]), to the upper ones ($\approx$ 
32, 41, 51 meV) as $x$ increases from $\approx$1/8 to the metallic state. 

As first proposed in \cite{Hennion3}, {\it the lower} levels can be interpreted in 
terms of quantized spin waves within $2D$ $F$ clusters of $4a$ size, considered 
as residual effects of the orbital ordering. For x$\approx$1/8, a peculiar 
ordering of these clusters is stabilized at low $T$ forming a ferromagnetic 
striped phase. {\it The upper ones} which have the main intensity in the metallic state, are assigned to their hybridization with the surrounding $3D$ $F$ matrix. By defining $J_{OD}$ 
and J$_{OO}$ as the nearest neighbours couplings in the matrix and 
the clusters respectively, we find that $J_{OD}$ increases with 
doping while J$_{OO}$ does not, so that $J_{OD} \le J_{OO}$ in zone 2 and 
$J_{OD} \ge J_{OO}$ in zone 4. As $J_{OD}$ is a measure of the stiffness 
constant $\cal D$ and in turn of the metallic coupling, this rapid variation 
with $x$ or $T$ is quite natural. Moreover, it explains the apparent 
softening of the spin wave spectrum\cite{Endoh,Ye,Moussa2}.

\begin{figure}[t]
\centerline{
\includegraphics[width=7 cm]{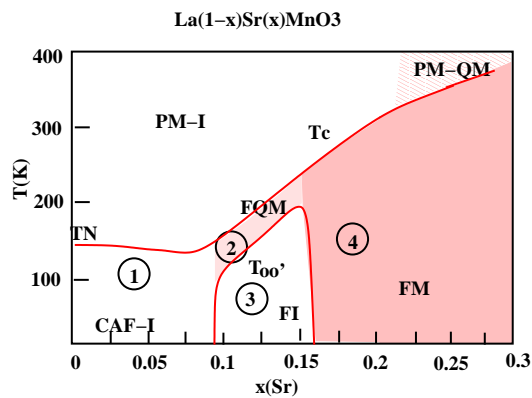}
}
\caption{(Color online) Schematic phase diagram of $La_{1-x}Sr_xMnO_3$. 
Zone 1: insulating canted antiferromagnetic state. 
Zone 2: ferromagnetic quasi-metallic state. 
Zone 3: insulating ferromagnetic state. 
Zone 4 : metallic ferromagnetic state.
}
\label{fig1}
\end{figure}

Inelastic neutron scattering measurements have been carried out at the 
Laboratoire L\'eon Brillouin ($1T, 2T, 4F$ spectrometers) and at the 
Institut Laue Langevin ($IN8$ spectrometer). To obtain a high resolution 
even at large energy transfer, $Cu_{111}$, $Cu_{002}$ or $Cu_{220}$ 
monochromators have been used. All samples have an orthorhombic 
structure except $La_{0.7}Sr_{0.3}MnO_3$ which is rhombohedral. For simplicity, 
the wavevector $q$ is defined in pseudo-cubic indexation with {\bf a, b, c} 
directions and $a$ as lattice spacing. In the quasi-metallic and insulating 
states of $La_{0.85}Sr_{0.15}MnO_3$ where the magnetic coupling is anisotropic, 
we consider that, by continuity with $LaMnO_3$, {\bf a} and {\bf b} are equivalent 
whereas {\bf c} is distinct. There, due to twinning, the [100], [010], [001] 
directions are superposed whereas the [111] direction corresponds to a 
single domain.

\begin{figure}[t]
\centerline{
\includegraphics[width=9 cm]{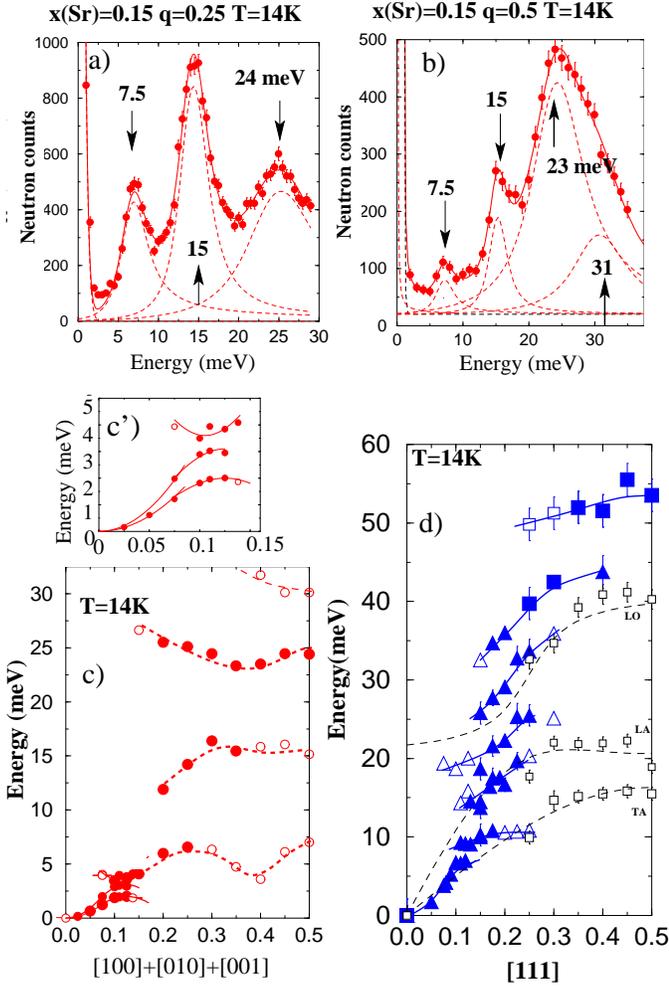}
}
\caption{(Color online) Spin waves determination in $La_{0.85}Sr_{0.15}MnO_3$ 
at 14K. a) and b): examples of energy spectra at Q=(1.25,0,0) and (1.5,0,0) 
resp., c) and d): spin wave spectra along [100]+[010]+[001] and [111] resp. 
with, in (c'), a zoom of the low energy part of Fig 1-c. The full (empty) color 
of the symbols indicates large (weak) intensity. In d), triangles (squares) indicate measurements in second (third) Brillouin zones. The black squares are phonons measured with magnons in the third Brillouin zone. 
}
\label{fig2}
\end{figure}

\begin{figure}[t]
\centerline{
\includegraphics[width=9 cm]{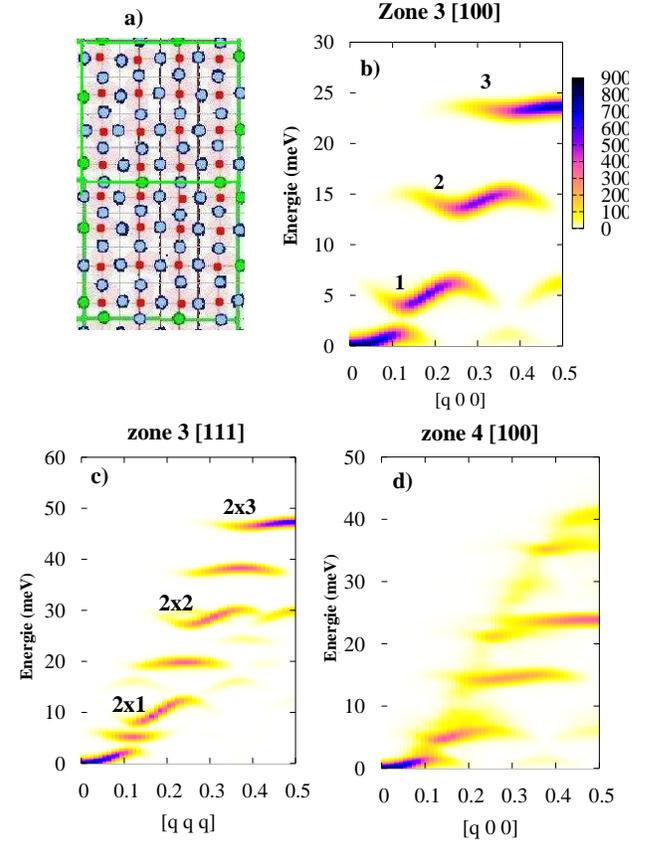}
}

\caption{(Color online) {\bf a}): schematic drawing of two units of the 2D stripe model with red (blue and green) circles for Mn (O). {\bf b, c, d}): calculated spin wave spectra (see the text). {\bf b}): zone 3, along [100], {\bf c}) zone 3, along [111], {\bf d}): zone 4, along [100] using $44 \times 44$ spins in ({\bf a, b}) plane and 5 planes with cyclic boundary conditions. 
}
\label{fig3}
\end{figure}

\begin{figure}[t]
\centerline{
\includegraphics[width=7.5 cm]{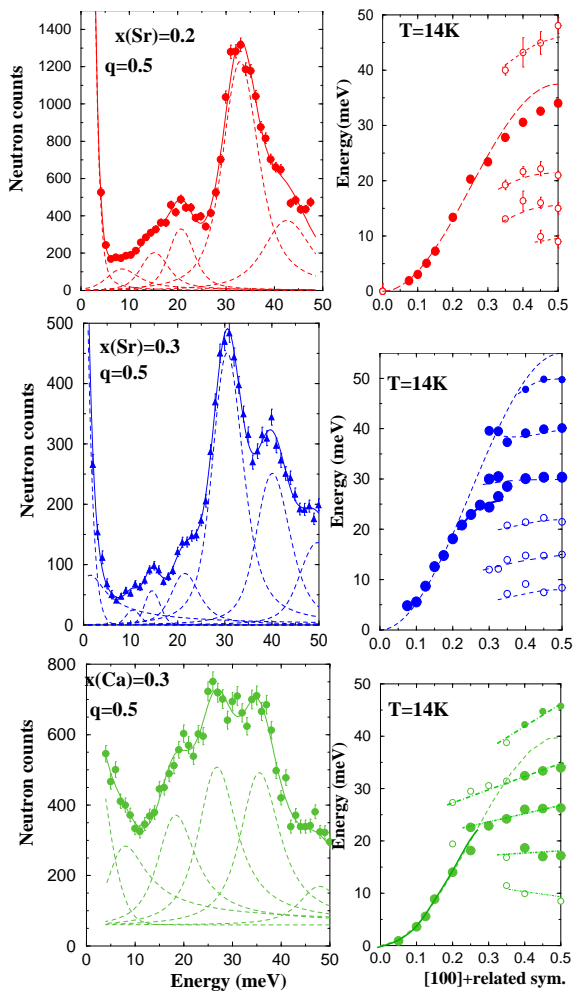}
}
\caption{(Color online) Raw data at Q=(1.5,0,0) (left panel) and spin wave 
dispersions (right panel) determined at 14K for $La_{0.8}Sr_{0.2}MnO_3$ (top), 
$La_{0.7}Sr_{0.3}MnO_3$ (middle) and $La_{0.7}Ca_{0.3}MnO_3$ (bottom). The full (empty) color corresponds 
to large (weak) intensity. The dashed line points out the softening effect. Note the change of scale with 
$x(Sr)$=0.15, Fig 2.
}
\label{fig4}
\end{figure}

\begin{figure}[t]
\centerline{
\includegraphics[width=7 cm]{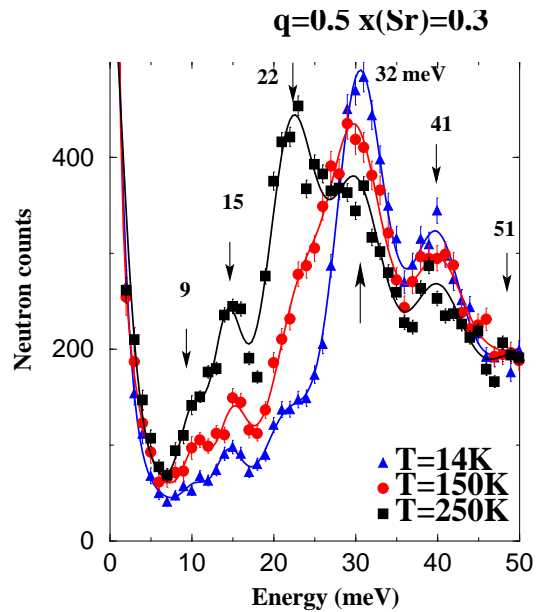}
}
\caption{(Color online) Raw data for Q=(0,0,1.5) 
in $La_{0.7}Sr_{0.3}MnO_3$ at T=14K (blue triangle), 150K (red circle) and 
250K (black square). 
}
\label{fig5}
\end{figure}

We first describe the spin dynamics in zones $2$ and $3$ of the phase diagram.
Here, the spin wave spectrum of $x(Sr)=0.15$ along [1+q,0,0] is found to be 
very close to that observed for $x(Sr)=0.125$ \cite{Hennion3}. Below T$_C$=230K, 
a quadratic $E=\cal D$$q^2$ dispersion is observed 
for $q \le 0.25$, as well as several q-independent levels for $q>0.25$. Below T$_{OO'}$=180K, 
$\cal D$ increases, whereas a gap opens at $q$=0.125. Concomitantly, the 
levels become $q$-modulated. Examples of raw data are shown at $T$=14K in Fig 
2-a and Fig 2-b. The arrows point three resolved levels at $E$=7.5, 15, 23 meV 
for q=0.5. The mode at $E\approx 31$meV in the shoulder of the main peak will be  discussed 
below. The spin wave spectrum is shown in Fig 2-c along [100] and in Fig 2-d 
along [111]. The main energy values at zone boundary along [100] and [111] 
differ by a factor of $\approx$ 2, which is expected for a $2D$ coupling. 
Actually, thanks to the high-energy resolution, the data provide evidence 
for two new features, not observed before \cite{Moussa1}. 

i) Along [1+q,0,0] with $q$=$0.125$, three modes are detected (cf Fig 2-c 
and 2-c'). The curve with a lower enery, observed only for $q<0.15$, is assigned, 
from its $T$ variation, to the {\bf c} direction. As a result, both {\bf a} 
and {\bf b} directions are concerned with the gap opening.

ii) Along [111], six levels are resolved beyond the small-q dispersion (Fig 2-d). 

In a recent paper \cite{Hennion3}, we proposed a qualitative interpretation of 
this spin wave spectrum in terms of quantized spin waves in $2D$ $4a \times 4a$ 
ferromagnetic clusters. Aiming to improve this description, we propose a 
phenomenological model, assuming the existence of a long-range ordering 
of such clusters in ({\bf a, b}) planes, resulting in the stripe picture sketched in Fig 3-a. In 
absence of any experimental indication, we do not consider any ordering 
along {\bf c}. We then use a Heisenberg model with two next nearest neighbors 
($NN$) coupling constants: $J_{OO}$ and $J_{\mbox{inter}}$ couple respectively 
Mn spins within the same clusters and across the cluster boundaries. This 
superstructure, also proposed in the context of high-T$_C$ \cite{Boris}, implies that the holes lie, in average on the oxygen ($O$) atoms of the boundaries so that x$=1/8$ precisely corresponds to half-filling (see Fig 3-a). This can explain its stability on a large $x$ range.

 In this picture, the spin 
dynamics is that of quantized spin waves within $4a\times 4a$ clusters. 16 levels 
are expected, whose energies are given by 
$E_{x,y}=4J_{OO}S \left (1-\frac{\cos{\frac{\pi x}{n}}+\cos{\frac{\pi y}{n}}}{2}\right)$, 
with $n=4$ and $x,y=0,..3$.
Choosing $J_{\mbox{inter}}$=$J_{OO}/5$ allows a {\it quantitative} description of 
the data, at least for $q>$0.15 (fig 3-b, Fig 3-c). Because of the cluster square symmetry, three levels along [100] and six along [111] can be observed, in good agreement 
with experiment. The discrepancies which appear for $q<$0.15 along [100] (too 
large gap at $q=0.125$) and along [111] (calculated energy values smaller than 
observed) both indicate that another F component coexists, giving rise to the 
nearly isotropic coupling seen for q$<$0.15 (cf Fig 2-c'), or, equivalently, that 
the $2D$ ordering of the clusters is not very long-range \cite{Papavassiliou}. Before going further, 
we point out that our confidence in this model arises from the agreement found for both the [100] and [111] directions. Other models 
have been built by considering structural observations such as the (0,0,1/4) 
peak\cite{Yamada, Geck}. Actually, a similar spin dynamics is observed in all 
the compounds which exhibit the same anomaly of resistivity (zone 3), whereas, the structural 
anomalies differ. An incommensurate superstructure peak is found for x(Sr)=0.15, whereas for x(Ca)=0.17 and x(Ba)=0.15 short-range structural defects are 
observed (\cite{Fernandez-Baca} and to be published).

This new quantitative analysis of the spin wave spectrum allows to go a step 
further, to the true metallic state.  
For $x$ at and beyond the insulator-metallic transition, the 
experimental observations show strong similarities with the 
$x\approx 1/8$ range. 
In fig 4, left panels show the raw data for $x(Sr)=0.2$ (T$_C$=325K) 
(similar to $x(Sr)=0.175$), $x(Sr)=0.3$ (T$_C$=370K) and $x(Ca)=0.3$ (T$_C$=255K) 
at $q$=0.5, $T$=14K. Right panels show the corresponding dispersions. 
Again, the spectra consist of two components : a quadratic regime near 
the zone center (with a weak modulated intensity around) and $2$ (possibly $3$) levels, namely at $15$ and $22$ meV, as observed in the x$\approx$1/8 
range. 
Additional ones are observed at higher energies, namely $E$=32 meV, 
$\approx$ $41$ meV, $\approx$ $51$ meV for 
$x(Sr)$=0.3, and very close values for $x(Ca)$=0.3 considering the slight q dependence of the levels.

Whereas the energy values of the levels are very close in all samples, 
their intensity shows a remarkable evolution, that we shall correlate 
with the evolution of the stiffness constant $\cal D$. A large $\cal D$ value is associated with a strong intensity on the highest energy levels. This 
can be easily demonstrated by comparing $x(Sr)=0.3$ and $x(Ca)=0.3$ 
($\cal D$=200 and 150 meV $\AA^2$ respectively) in Fig 4. Similarly, as $\cal D$ decreases with increasing temperature, the intensity of the low-energy 
levels increases, showing again the inter-dependence of the 
two components. Such a variation is displayed in Fig 5 for $x(Sr)=0.3$, 
where at constant energy levels, the intensity balances from upper to 
lower levels as $T$ increases from 14K to 250K. The levels persist 
above T$_C$. 

Along [111], new levels are observed above $50$ meV up to $\approx 90-100$ meV,
which reveal a $3D$ coupling. This will be reported elsewhere.

These observations in the metallic state can be understood in the framework 
of the model discussed above. We consider an extended version, assuming 
disordered $2D$ $F$ clusters of $4a\times 4a$ size, embedded in a $3D$ 
$F$ matrix. Here, an additional ferromagnetic $NN$ coupling, $J_{OD}$ 
acting between spins within the $3D$ medium is introduced. A qualitative  agreement 
between the experiment and numerical simulations is obtained if $J_{OD}>J_{OO}$. 
Fig 3-d shows an example of calculated spin wave spectrum along [100] performed with a 
density of $50$\% of clusters, $J_{OD}=1.75J_{OO}$ ($J_{OO}$=1.7 meV) and 
$J_{\mbox{inter}}$=J$_{OO}/5$. The levels reminiscent of quantized spin 
waves within clusters are still there. However, the hybridization with 
the $3D$ $F$ medium results in blurring the large gaps at low $q$ and by 
adding new levels at large $q$. Actually, the weak $E$=31 meV level detected 
in the $x\approx$ 1/8 range below T$_{OO'}$ (Fig 2-b, 2-c and \cite{Moussa1}) 
could be also due to this effect, in agreement with NMR spectroscopy \cite{Papavassiliou}. Considering now the 
whole studied doping range, J$_{OO}$, which can be associated with superexchange (coupling induced by bound electrons)
is found to remain approximately constant. In contrast $J_{OD}$, which has 
basically the same meaning as $\cal D$ ($J_{OD}$=$\cal D$$/2Sa^2$) (coupling induced by hopping electrons) is found to increase 
with doping. 

Our experiment provides a simple explanation for the so-called "softening" effect: the zone boundary energy value determined on the basis of a cosine law from the $q\approx$0 dispersion (dashed lines in the right 
panels of Fig 4), clearly lies above the experimental 
one. This description of the data assumes however the existence of a 
single branch\cite{Furukawa}. In our view, this is however not the case, as the zone 
boundary unravels the quantized levels. The apparent "softenig" is simply 
due to the fact that $J_{OD} \ge J_{OO}$. Note that our interpretation reproduces 
the experiments in zone 2 with $J_{OD} \le J_{OO}$.

In conclusion, instead of the previous analysis with a fourth neighbor coupling\cite{Endoh,Ye,Moussa2}, the new data allow to describe 
the metallic state in terms of quantized spin waves in $2D$ clusters embedded in a $3D$ matrix. The remarkable similarities of the spin dynamics observed in several compounds, indicate that these inhomogeneities have the same origin 
whatever $T$, $x$ or the average cation size. 
These new observations should be important to understand the residual 
resistivity \cite{Viret} and the distinct magnetic time scales far from T$_C$\cite{Heffner}. Moreover, it provides a new insight around T$_C$ where magnetic anomalies are observed \cite{Lynn,Vassiliu}, and, hence, in the giant magnetoresistance, qualitatively similar in compounds with $x(Ca)$ and $x(Sr)$ ($x(Sr)$$<$0.4)\cite{Schiffer,Urushibara}, even if the "correlated polarons" play a role in Ca substituted samples \cite{Kiryukhin}. In addition, this study leads to a unified picture from the CAF state with hole-rich 
platelets \cite{Hennion1} to the F metallic state with hole-poor ones.

\begin{acknowledgments}

\end{acknowledgments}

\end{document}